\documentclass[a4paper]{article}

\usepackage{icrc2013}
\usepackage{booktabs}
\usepackage{url}
\usepackage{amssymb}
\usepackage[UKenglish]{babel}

\title{A large-area single photon sensor employing wavelength-shifting and light-guiding technology}

\shorttitle{Wavelength-Shifting Single Photon Sensor}

\authors{
Lukas Schulte$^{1}$,
Markus Voge$^{1}$,
Akos Hoffmann$^1$,
Sebastian B\"{o}ser$^{1}$,
Lutz K\"{o}pke$^{2}$,
Marek Kowalski$^{1}$
}

\afiliations{
$^1$Physikalisches Institut, Universit\"{a}t Bonn, Nu\ss{}allee 12, D-53115 Bonn, Germany\\
$^2$Institut f\"{u}r Physik, Universit\"{a}t Mainz, D-55099 Mainz, Germany
}

\email{schulte@physik.uni-bonn.de}

\abstract{Large-scale underground water-Cherenkov neutrino observatories rely on single photon sensors whose sensitive
area for Cherenkov photons one wants to maximise. Low dark noise rates and dense module spacing will thereby allow
to substantially decrease the energy threshold in future projects.
We describe a feasibility study of a novel type of single photon sensor that employs organic wavelength-shifting
material (WLS) to capture Cherenkov photons and guide them to a PMT readout.
Different WLS materials have been tested in lab measurements as candidates for use in such a sensor and photon
capture efficiencies as high as 50\,\% have been achieved. Based on these findings we estimate that the effective
photosensitive area of a prototype built with existing technology can easily exceed that of modules currently used e. g.
in IceCube. Additionally, the dark noise rate of such a module can be exceptionally low in the order of 10\,Hz. This is
of special importance when targeting low-energy neutrinos that yield only few photons that need to be distinguished from
noise hits.}

\keywords{water cherenkov, photon sensor, wavelength-shifter, low noise.}

\begin{document}
\maketitle


\section{Motivation}

Current large-scale water Cherenkov detectors in neutrino physics---such as IceCube \cite{bib:icecube} or
SuperKamiokande \cite{bib:superk}---instrument large amounts of water or ice with photosensitive devices. Efficient
collection of the Cherenkov light is usually achieved by using large numbers of large photomultiplier tubes (PMTs). 
The fluxes and energies accessible by the experiment are then determined by the total mass of instrumented material,
the density of instrumentation and the photo-collection area of each sensor. For IceCube, more than 5000 10'' PMTs
have been deployed into one km$^3$ of ice, which detect a sufficient number of photons for neutrinos exceeding an energy
of 100\,GeV.

When one intends to lower the energy threshold significantly while keeping the effective mass at the megaton-level
to target e.\ g.\ MeV neutrinos from extragalactic supernovae, two problems arise. Firstly, the noise rate of
PMTs is proportional to their cathode area, so at some point a low-energy event that only has a low photon multiplicity
can easily be mimicked by a random coincidence of module noise hits \cite{bib:micaSN}. Secondly, covering an effective
photosensitive area exceeding that of current experiments by factors of tens with conventional PMT technology is a
significant (financial) challenge.

In the following we will describe the Wavelength-shifting Optical Module (WOM), a single photon sensor that
can be used in extreme environments such as the deep antarctic glacial ice and has a total sensitive area exceeding that
of current technology sensors by a factor of a few while its noise rate is reduced by up to two orders of magnitude. 
While the potential of the technology is manyfold, we concentrate our efforts on the future extensions of IceCube.

\section{The wavelength-shifting optical module}
\label{sec:WOM}

 \begin{figure}[t]
  \centering
  \includegraphics[width=0.5\textwidth]{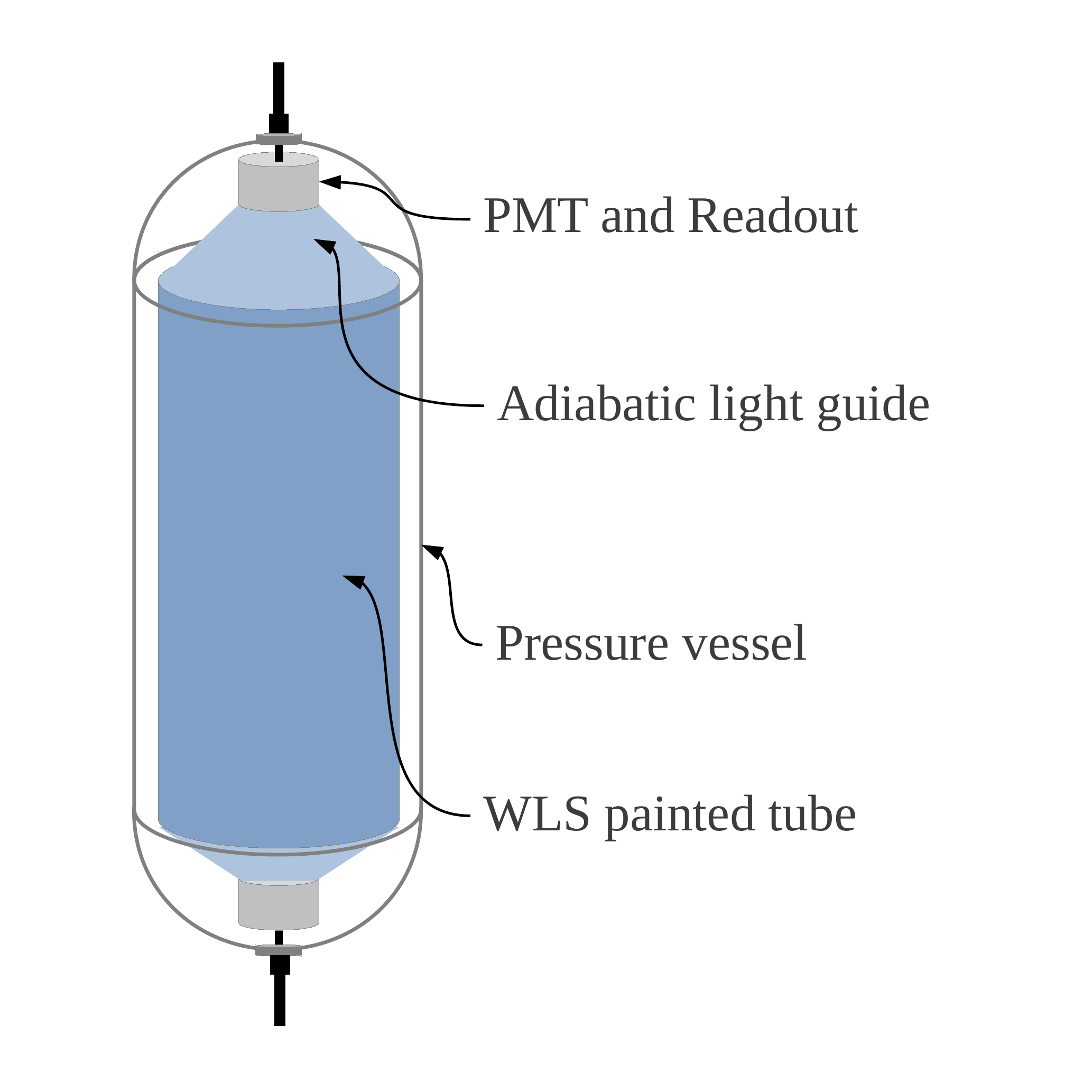}
  \caption{Design concept of the Wavelength-shifting Optical Module (WOM).}
  \label{fig:WOM_design}
 \end{figure}

The main idea of the WOM is to increase the sensitive area of a PMT by using passive components that act as light
collectors and concentrators (a sketch of the module design is shown in Fig.~\ref{fig:WOM_design}). Due to Liouville's
theorem, mirrors or lenses can not be used for this purpose, however, by introducing a wavelength shift (at the cost of
a short time delay), this limitation can be overcome.

We envision cylindrical tubes that have wavelength-shifting properties. These tubes collect Cherenkov photons on
their outer surface. These Cherenkov photons, which are mainly in the UV regime, are absorbed and then re-emitted
isotropically at a larger wavelength within the tube. The isotropic emission ensures that a large fraction 
of the photons, which were incident roughly perpendicular to the surface, will now be captured inside the tube
and then guided towards the end via multiple total internal reflection.

At both ends of the tube, small high-efficiency PMTs will be placed that read out the incoming photons. Since their
spectrum has been shifted away from UV towards the optical blue, it is now better suited for readout by conventional
PMTs as those are usually most sensitive in the optical blue and green.

The whole assembly will be enclosed by a transparent pressure vessel that protects the components from physical damage.
It will also ensure that the wavelength-shifting tube is surrounded by a small gap of air and not in direct optical
contact with the outside (e.\,g.\ the glacial ice). Although an optical contact to the surroundings usually is
desirable to gather as many photons as possible, in our case a large difference in refractive index is needed to
achieve efficient total internal reflection.

The three most important components of the WOM, for which we will discuss details of their properties, are:

\begin{itemize}
 \item the pressure housing
 \item the PMT
 \item and the wavelength-shifting tube.
\end{itemize}

For the pressure housing, apart from mechanical stability\footnote{Static pressure of up to 10000 psi can occur during
the freeze-in process when deploying in deep antarctic ice.}, two requirements have to be met. Firstly, optical
transparency has to be guaranteed not only in the optical but also in the UV regime down to $\approx$\,250\,nm.
Secondly, the material needs to have a high radio purity. In regular glass, $^{40}$K is a common contamination and its
decay would give rise to a very high module noise rate. It turns out that fused quartz glass is a very good
choice, having a transparency of 90\,\% at 250\,nm and an activity of only 0.02\,Bq/kg \cite{bib:quartzprops}.

As a possible PMT we consider a prototype that has an novel photocathode with enhanced green sensitivity
\cite{bib:hamamatsu}. Although for this prototype no noise measurements at low temperature are available, for
similar models typical noise rates are reported to be below 1\,Hz per cm$^2$ of cathode area at a temperature of
--30\,$^\circ$C
\cite{bib:meyerPMT}.

The wavelength-shifting tube can be made of quartz glass as well although the wall thickness can be much thinner, since
there is no pressure inside. This inner cylinder will be covered by a thin film of wavelength-shifting material.
First tests have been done with samples produced by dip-coating 20\,mm and 5\,mm diameter glass tubes with EJ-298
wavelength-shifting paint, which consists of a fluorescent dopant and a PVT base \cite{bib:paintprops}. The film
thickness is $\approx$\,50\,$\mu$m, its surface RMS has been measured with an AFM to be 3\,nm.
The wavelength shifter itself is an organic material with an inherent activity of 0.4\,Bq/kg \cite{bib:pvtnoise}, yet
its contribution is negligible due to the thinness of the layer. Hence the noise rate of this component is dominated by
the quartz glass tube. 

In the following section, we discuss the efficiency of the different components.

\section{Efficiency calculation}

\subsection{Photon capture efficiency}
\label{sec:capture_eff}

 \begin{figure}[t]
  \centering
  \includegraphics[width=0.5\textwidth]{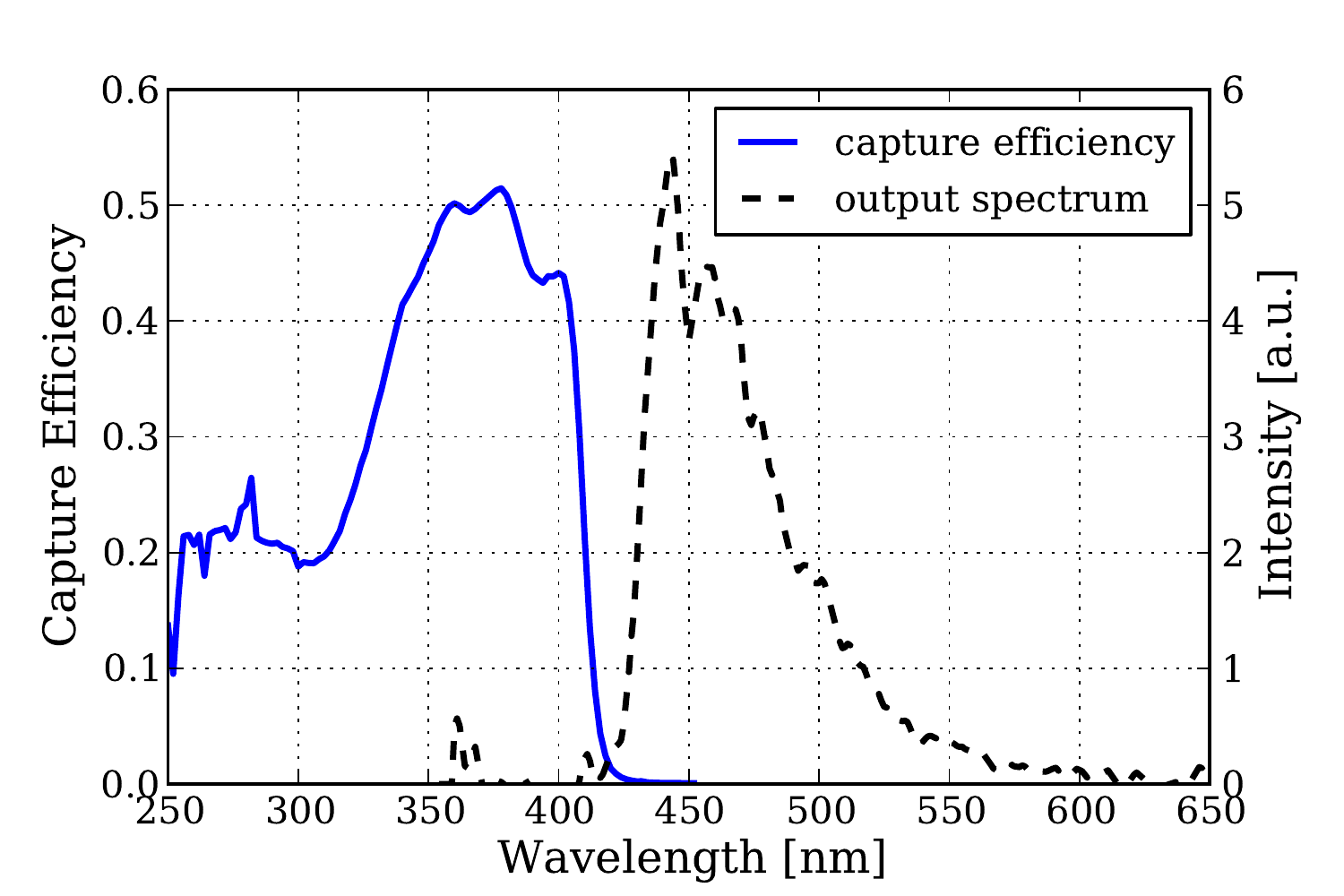}
  \caption{Capture efficiency and output spectrum of a 20\,mm diameter fused quartz glass tube coated with EJ-298
           wavelength-shifting paint.}
  \label{fig:capture_eff}
 \end{figure}

The photon capture efficiency ($CE$), discussed in the following paragraphs, is defined as
\begin{equation}
 CE(\lambda) = \frac{\mathrm{\sharp\ detectable\ photons}}{\mathrm{\sharp\ photons\ injected\ at\
outer\ surface}}\quad.
\end{equation}

We have measured this quantity in a lab setup using a monochromator with wavelength range 250\,--\,1200\,nm and two
identical photo-sensors, read out via lock-in amplifiers.
The emission spectrum and capture effiency as a function of wavelength are shown in Fig.~\ref{fig:capture_eff} for the
20\,mm sample described in the previous section. For the 5\,mm sample, the results are identical within the precision of
our measurement.

The peak efficiency of $\approx$\,50\,\% has to be compared to the theoretical maximum, which, assuming that all
incoming photons are absorbed by the active dye and then re-emitted isotropically with 100\,\% quantum effiency, and
ignoring any transport loss, is given by the fraction of solid angle where total internal reflection occurs:
\begin{equation}
 CE_{\mathrm{max}} = \cos(\vartheta_c) = \cos(\arcsin(\frac{n_\mathrm{air}}{n_\mathrm{glass}})) \approx 75\,\%\quad.
\end{equation}

The decrease in efficiency down to $\approx$\,20\,\% below 350\,nm is assumed to be due to reabsorption and/or UV
absorption inside the PVT base of the paint. This should be reducible by thinning the paint layer and future R\&D for
the dye. An improved production scheme allowing for thinner layers is under development.

\subsection{PMT readout efficiency}

To account for the fact that the captured photons coming out at either end of the wavelength-shifting glass tube will
be read out by PMTs, its (normalized) output spectrum (also shown in Fig.~\ref{fig:capture_eff}) has to be convolved
with the quantum efficiency of a realistic PMT. This gives the fraction $\varepsilon_\mathrm{PMT}$ of photons coming out
of the wavelength-shifting tube that are actually detected by the PMT. Using the data provided by Hamamatsu for their
R7600-EG prototype \cite{bib:hamamatsu}, we arrive at a readout efficiency of $\varepsilon_\mathrm{PMT} = 30.8\,\%$.

\subsection{Angular efficiency}

\begin{figure}[bht]
  \centering
  \includegraphics[width=0.5\textwidth]{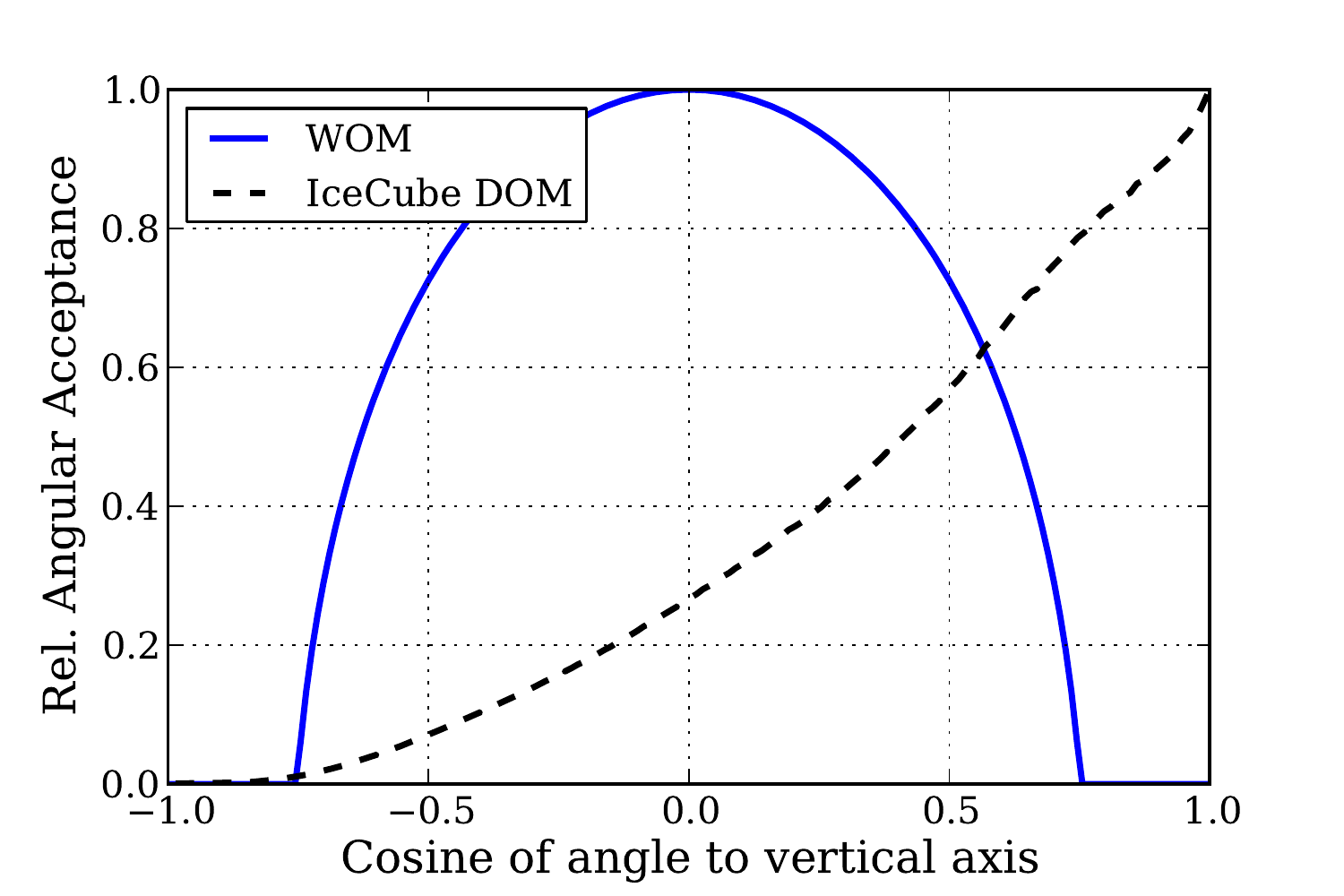}
  \caption{Relative angular acceptance of WOM and IceCube DOM.}
  \label{fig:angular_eff}
\end{figure}

The angular acceptance $\varepsilon_\Omega$ of the WOM can be calculated with Fresnel's equations. We assume a
homogeneous, parallel bundle of light incident onto the WOM at an angle $\vartheta$ w.r.t.\ its longitudinal (i.e.\
vertical) axis. The acceptance at this angle is calculated by integrating the transmission probability from ice
($n=1.33$) through glass ($n=1.48$) into air ($n=1.0$) over the whole visible surface of the WOM, including the fact
that the WOM appears shortened by a factor of $\sin(\vartheta)$ at angles different from $90^\circ$. 
Note that this only covers the propagation of photons from the surrounding ice into the air gap, since the penetration
from air into the actual wavelength-shifting tube is included in the capture efficiency described in
Sec.~\ref{sec:capture_eff}.

In Fig.~\ref{fig:angular_eff} we show the relative acceptance with respect to the maximum for the WOM and an IceCube
Digital Optical Module (DOM), the photosensor module used in IceCube \cite{bib:ICdom}. For the WOM, a maximum acceptance
$\varepsilon_\Omega(\mathrm{max}) = 70.1\,\%$ is achieved at $\vartheta=90^\circ$. Averaging the relative angular
acceptance shown in Fig.~\ref{fig:angular_eff} over all angles of incidence, the mean angular acceptances are
$\bar\varepsilon_\Omega(\mathrm{WOM}) = 57.5\,\%$ and $\bar\varepsilon_\Omega(\mathrm{DOM}) = 34.1\,\%$, respectively.

\subsection{Full module efficiency}

\begin{figure}[tbh]
  \centering
  \includegraphics[width=0.5\textwidth]{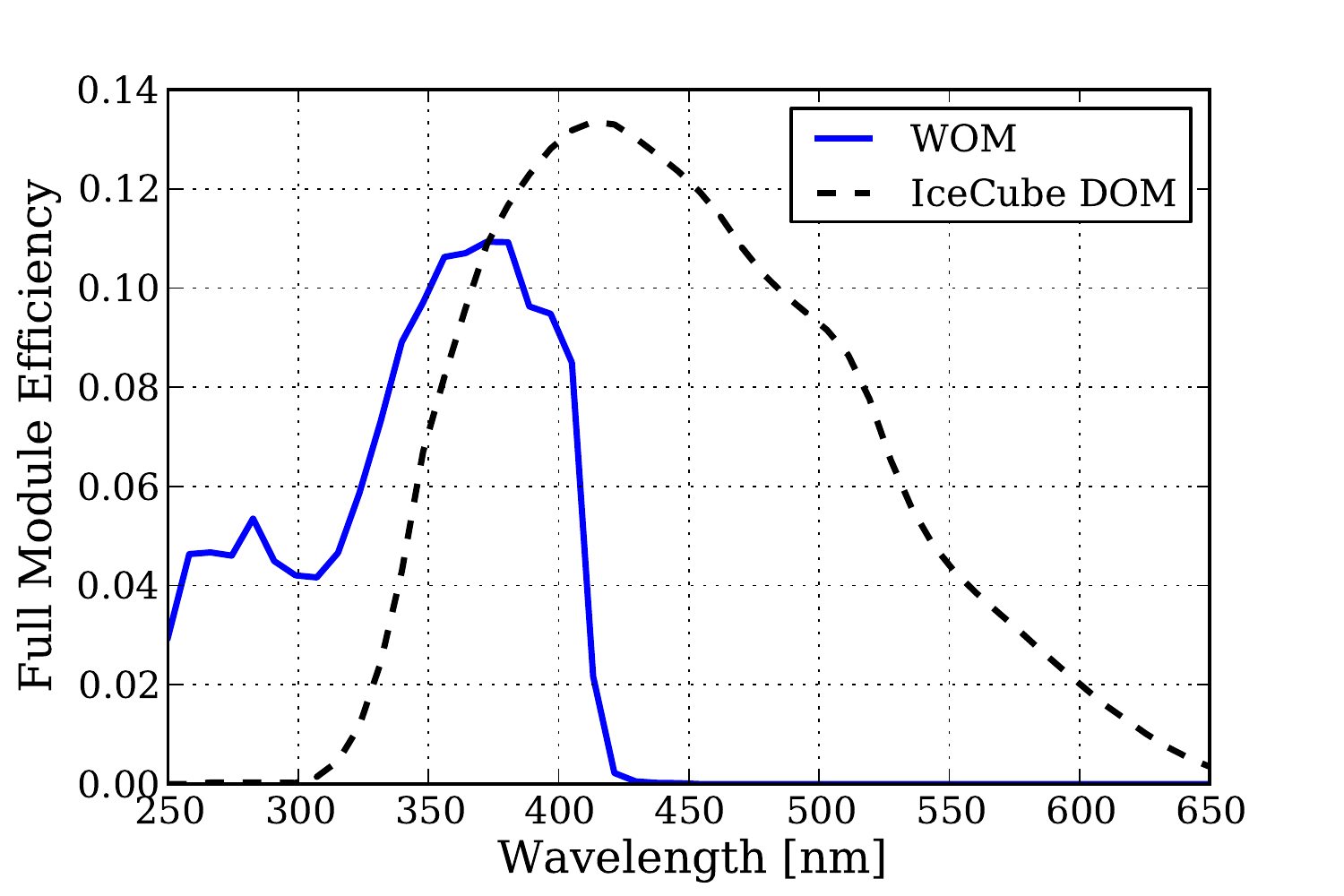}
  \caption{Full module photo-detection efficiency of WOM and IceCube DOM at optimal illumination angles.}
  \label{fig:full_eff}
\end{figure}

To obtain the detection efficiency of the fully assembled module, the three efficiencies discussed above,
$CE$, $\varepsilon_\mathrm{PMT}$, and $\varepsilon_\Omega$, have to be multiplied:
\begin{equation}
 ME(\lambda) = \varepsilon_\Omega(\mathrm{max}) \cdot \varepsilon_\mathrm{PMT} \cdot CE(\lambda)\quad.
\end{equation}

The result, assuming optimal illumination (i.e.\ $\vartheta=90^\circ$), is shown in
Fig.~\ref{fig:full_eff}, compared to the efficiency of an IceCube DOM at its optimal illumination angle
($\vartheta=0^\circ$).

\section{Other properties}

\subsection{Noise}
\label{sec:noise}
Two components of the WOM dominate its noise rate: the fused quartz glass tubes and the PMTs. Clearly, the
pressure housing will contribute much more than the painted tube in the interior simply due to its larger mass.

The required thickness depends strongly on the total diameter. Assuming a diameter of 20\,cm of the pressure
housing, a wall strength of $\approx$\,2\,cm should be sufficient. 
With the typical activity of 0.02\,Bq/kg stated in \cite{bib:quartzprops}, this translates to a noise rate of 1\,Hz per
m$^2$ of sensitive area.

To read out the WLS painted tube in the interior, its ring-shaped end faces need to be projected onto single PMTs using
adiabatic light guides, Winston cones or similar. Then a total PMT cathode area equal to the size of twice its end face
(since each end will be equipped with a PMT) is needed. Since the inner tube can have very thin walls, this area
should be at the order of tens of cm$^2$. According to the numbers from \cite{bib:hamamatsu, bib:meyerPMT}, this
corresponds to a PMT noise rate below 10\,Hz.

So assuming a WOM with a diameter of 20\,cm and a length of 2\,m, hence a sensitive area of $2\pi R L \approx
1.3\,\mathrm{m}^2$, a total module noise rate in the order of only 10\,Hz seems possible.

\subsection{Time Resolution}

Using a toy Monte Carlo, we have simulated the propagation of photons inside the wavelength-shifting tube. Since there
is no possibility to reconstruct the point at which a detected photon actually entered the WOM, the variance of the
full travel time distribution has to be considered for the time resolution of the module. This number is proportional
to the length of the module with a value of 1.4\,ns/m.

A similar contribution to the timing uncertainty arises from the fact that, in order to be detected, the photons have to
be absorbed and re-emitted by one of the wavelength-shifting dye molecules. The typical emission time is stated by
the manufacturer to be 2.0\,ns \cite{bib:paintprops}.

\section{Conclusion}

Comparing the module efficiencies of WOM and IceCube DOM in Figs.~\ref{fig:angular_eff} and \ref{fig:full_eff}, at first
sight the WOM only provides a moderate improvement in photon collection efficiency. However, one has to keep two things
in mind:

First, the WOM can be produced in a very large size since its size-critcal components are just two glass tubes (pressure
vessel and WLS painted inner tube), that are easily scalable. The size of the DOM on the other hand is given by the
size of the enclosed PMT which can hardly be pushed much further at reasonable cost\footnote{This is mostly due to the
deployment in the antarctic glacier: Larger spherical DOMs need larger diameter drill holes, dramatically increasing
the costs for drilling.}.

\begin{table}[tbh]
\begin{center}
\begin{tabular}{ccccc}
\toprule
Module & $\overline{ME}$ & $\bar\varepsilon_\Omega$ & Eff. Area     & Noise             \\ 
\midrule
WOM    & 4.40\,\%        & 57.5\,\%                 & 101\,cm$^2$   & $\approx$\,10\,Hz \\ 
DOM    & 5.36\,\%        & 34.1\,\%                 & 12.9\,cm$^2$  & 800\,Hz           \\
\bottomrule
\end{tabular}
\caption{Comparison of WOM and IceCube DOM properties for a Cherenkov spectrum between 250 and 600\,nm.}
\label{tab:comparison}
\end{center}
\end{table}

Second, the WOM is more sensitive in the UV below $\approx$\,370\,nm. As the photons that are to be detected stem from
Cherenkov radiation, their spectrum is proportional to $1/\lambda^2$. Accounting for this initial spectrum when
calculating the mean module efficiency
\begin{equation}
 \overline{ME} = 
\frac{ \int_{250\,\mathrm{nm}}^{600\,\mathrm{nm}} ME(\lambda) \frac{\mathrm{d}\lambda}{\lambda^2}}
{\int_{250\,\mathrm{nm}}^{600\,\mathrm{nm}} \frac{\mathrm{d}\lambda}{\lambda^2} }\quad,
\end{equation}
the performance of the WOM improves significantly w.\,r.\,t.\ the IceCube DOM.

Assuming a WOM with a diameter of 20\,cm and a length of 2\,m as in Sec.~\ref{sec:noise} and, for comparison, a DOM
with 30\,cm diameter \cite{bib:ICdom}, convolving both with a Cherenkov spectrum between 250 and 600\,nm, we can now
calculate the full effective area of WOM and DOM for an isotropic flux. The results are shown in
Tab.~\ref{tab:comparison}.

So in terms of total effective area, 
\begin{equation}
 A_\mathrm{eff} = \overline{ME}\cdot \bar\varepsilon_\Omega\cdot A_\mathrm{xsec}\quad,
\end{equation}
one WOM with a cross-section of $A_\mathrm{xsec} = 2 R L = 4000$\,cm$^2$ is the equivalent of about eight IceCube DOMs
($A_\mathrm{xsec} = 707$\,cm$^2$). If one would reduce the considered wavelength range to 300\,--\,600\,nm, the mean
module efficiency increases by about 40\,\% for the DOM (since its ``blind'' regime below 300\,nm does not enter in
the calculation), while for the WOM it remains essentially constant.


In summary, our proposed concept offers a technological solution to obtain a large area single photon sensor with
exceptionally low noise rate. Future work will concentrate on the optimization of the film coating, the choice of dyes,
as well as on the proof of long term stability and large scale expandability.

\vspace*{0.5cm}
\footnotesize{{\bf Acknowledgments:}{\ We thank 
Christian Weinheimer for fruitful discussions. This work has been supported by the Helmholtz Alliance for Astroparticle
Physics (HAP).}}

\end{document}